\documentclass[review]{elsarticle}

\usepackage{lineno,hyperref}
\usepackage{pxfonts} 
\modulolinenumbers[5]
\usepackage[symbol]{footmisc}

\usepackage{float}

\begin{document}
\begin{frontmatter}
\vspace{-50.0cm}
\title{DOBC-Based Frequency \& Voltage Regulation Strategy for PV-Diesel Hybrid Microgrid During Islanding Conditions}
\vspace{-5.0cm}
\author{Himanshu Grover\fnref{myfootnote}}
\author{Ashu Verma}
\author{T. S. Bhatti}
\address{Department of Energy Science and Engineering, Indian Institute of Technology Delhi, New Delhi, India}
\fntext[myfootnote]{Corresponding author.\\ Email address: himanshu.grover@iitd.ac.in (Himanshu Grover)}




\begin{abstract}
This paper proposes a disturbance observer-based control (DOBC) method for frequency and voltage regulation of a solar photovoltaic (PV)-diesel generator (DG) based hybrid microgrid during islanding conditions. The DOBC is integrated as a feed-forward control to the synchronous generator based DG, which handles real-time power mismatches and regulates the microgrid frequency and voltage under islanding. To substantiate the operational robustness of the developed controller under real-time uncertainties arising due to variability in PV output and load, the controller has been tested under worst-case uncertainty conditions. The proposed controller has been developed as a MATLAB/Simulink model and the results are validated on the real-time simulator OPAL-RT. The effectiveness of the proposed control scheme has further been validated in the presence of communication delays and noisy load conditions. Results verify the dynamic performance of the controller in regulating the system frequency and voltage for low-inertia microgrids. Finally, the proposed control strategy has been implemented on laboratory scale microgrid setup in which synchronous generator based diesel generator regulates system frequency fast and efficiently under worst case uncertainty scenario.  
\end{abstract}

\begin{keyword}
\texttt Disturbance observer-based control \sep solar PV \sep diesel generator \sep low-inertia microgrid \sep worst case uncertainty \sep communication delay  \sep noisy load.
\end{keyword}

\end{frontmatter}

\section{Introduction}
\begin{figure}[h]\centering
\includegraphics[width=4.5in]{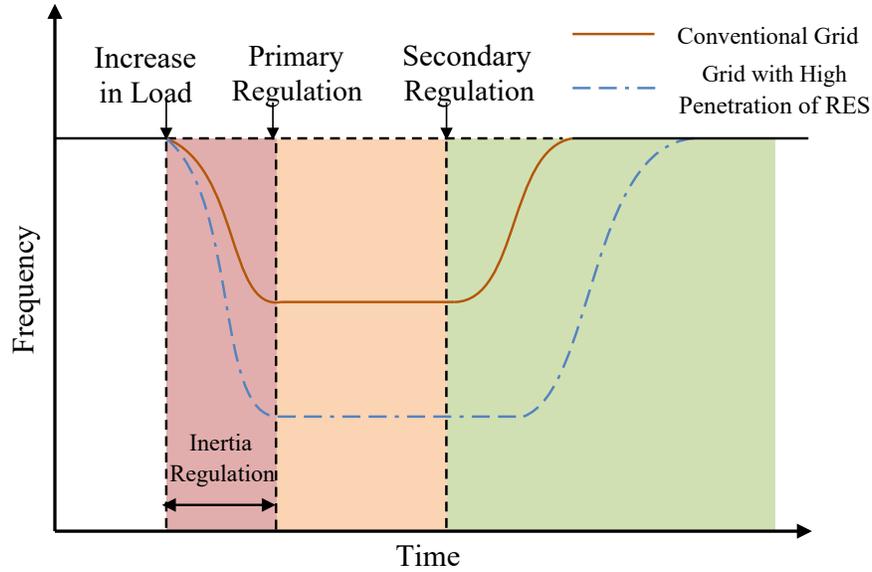}
\caption{A comparison of the frequency response of a traditional grid versus a grid and grid with high penetration of RES.}
\label{freq}
\end{figure}

Growing global concerns over carbon emissions and climate change have resulted in greater attention towards distributed energy systems such as microgrids which facilitate increased incorporation of renewable energy sources (RES) including solar photovoltaic (PV) and wind turbine generators (WTG). However, RES-based systems are often faced with reliability issues owing to their inherent variable and intermittent characteristics.
Furthermore, since RES have either low or zero inertia, integration of these generation sources results in greater challenges for the microgrid controller, in terms of system operation. It is noteworthy that the microgrid system inertia plays an important role for system stability during intermittency caused due to varying RES outputs and load. Large-scale integration of distributed RES results in low inertia of the microgrid, which causes large variations in system frequency and voltage due to power unbalance in the system \cite{REZAEI2015287}.
Figure \ref{freq} \cite{LIU2018169}, illustrates a comparison of the frequency response to a change in system power between a conventional grid and a grid with renewable energy penetration.  \\
\indent While the utility grid connection supports microgrid operation during normal operating conditions, a major challenge for renewable energy based microgrids is to provide frequency and voltage regulation during islanding condition. As the output of solar PV  depends upon the amount of solar insolation level falling over it and its geographical location, variation in solar PV output power causes large frequency and voltage deviation \cite{486595,1597338}. To handle such variability and intermittency, a stand-by energy source is required which handles energy demand during islanding condition. Diesel generators (DGs) and energy storage systems are often found to be a prominent backup source during islanding conditions, which reduces real-time load generation mismatch. A close review of relevant literature reveals several studies that have been conducted to minimize fluctuations caused by solar PV, using energy storage systems. 
Accordingly, the authors in \cite{5677458} proposes a fuzzy logic based control for a three-phase isolated solar PV-diesel microgrid, in which solar PV regulates the load voltage, whereas DG regulates the load frequency. Additionally, the authors in \cite{8804864}, proposed an approach for emergency frequency control for an isolated microgrid based on DG, solar PV, and battery energy storage systems.
Another work in this context was presented in \cite{ROSINI2021106974}, where the authors describe a model predictive control-based decentralised and communication-free technique for frequency and voltage management in a solar PV-energy storage-based microgrid. Authors in \cite{HIRASE2018699}, develops an algebraic-type based control approach for virtual synchronous generator (VSG), which reduces the frequency and voltage deviations in the system using battery energy storage system (BESS). It is observed from the literature that even through energy storage systems help in frequency regulation, however, their capital, maintenance and replacement costs increase the overall system cost. This is indicative of the fact that employing energy storage devices for minimizing fluctuations caused by RES during islanding conditions is not an economically viable solution for grid-connected microgrids.\\ 
\indent As a reliable power backup solution, most practical microgrids are observed to employ DG to provide reliability and energy security during islanding/isolated conditions. 
Due to their wide range of availability and controllability, DGs can better handle the fast fluctuating power outputs of solar PV, thereby, regulating the voltage and frequency of the islanded microgrid within specified limits. The research works presented in \cite{9113059}, \cite{AZIZ2022122458}, \cite{LIU2021125733} and \cite{KUMAR2021102965} demonstrate existing microgrids integrated with DG to increase the supply reliability in isolated and offgrid systems. Similarly, \cite{MARQUSEE2020114918,MARQUSEE2021116437} highlight the strategic importance of DG for meeting critical loads such as schools, hospitals etc. which demand uninterrupted power supply during a grid outage. Accordingly, a number of recent research works have focused on developing DG controls for optimal operation of microgrids. The authors in [15], propose a unified control architecture for a solar photovoltaic-diesel microgrid that runs in both grid-connected and islanded modes.
Furthermore, the authors in \cite{5648756} proposed a modified fuzzy logic based control scheme for solar PV-diesel microgrid without energy storage. Similarly, the authors in \cite{RASHED2008949} have developed a control strategy for a PV-diesel autonomous power system which regulates the voltage using interfacing inverter and a robust fuzzy logic controller is developed for regulating system frequency through DG. While the fuzzy logic control based approaches improve system performance of conventional controllers, their dynamic performance is highly dependent on the expert systems thereby, affecting the accuracy of the solution. Furthermore, these system from the \textit{curse of dimensionality} which increases the computational burden significantly as the dimension of the system increases. \\
\indent In this context the disturbance observer based control (DOBC) has been observed to enhance controller performance in relevant literature. 
A coordinated control strategy has been developed in \cite{8513881} to adjust frequency and voltage using adaptive sliding mode method with disturbance observer control. Similarly disturbance observer and double sliding mode controller strategy for frequency regulation has been proposed for a hybrid isolated microgrid \cite{7478160}. Recently, the authors in \cite{8676274}, presented DOBC control strategy in which VSC connected to energy storage regulates the frequency for a low inertia microgrid while DG runs in constant power mode. DOBC is seen to efficiently estimate any disturbances occurring in the system and to significantly increase system performance without imposing a significant computational burden.\\
\indent A number of research works have focused upon mitigating the effects of unforeseen uncertainties in generation and load forecasts. In view of this, \cite{9224611} explains different methods for handling uncertainties and mitigation methods. While a major segment of these works focus upon precise modeling of uncertainties and developing intelligent computational methods \cite{8630730,7764193,Uncertainty,ABEDINI2019100200,LANKESHWARA2022117971}, only few works have focused upon developing control techniques to handle uncertainties occurring during real-time system operation. A majority of such research works revolve around energy storage systems such as batteries and supercapacitors \cite{8283546,6894151,314514,ABUBAKR2021106814}, which in turn, significantly increase systems costs and are not economically viable solutions for small and remote microgrid systems. A few research works develop robust control techniques, wherein the controller explicitly focuses on ensuring reliability of controllers and their robustness to withstand uncertainties in the plant model parameters. However, this does not ensure operational robustness to real-time variations in generation and load, which are inherently uncertain and challenging to precisely model. Thus, it is imperative to develop controls that perform suitably well in real-time conditions and demonstrate practical robustness. A major challenge in this context is to identify the uncertainty conditions under which the developed controls can be practically tested and verified for robustness. In view of this discussion, this paper develops robust uncertainty conditions to test the control technique and evaluate its real-time performance. The robust test conditions are obtained through maximization of uncertainty occurrence, as worst-case uncertainties using well-defined polyhedral uncertainty sets.\\
\indent Keeping in view the issues in coordinating the operation of solar PV and DG during islanding conditions for a low-inertia microgrid, a robust controller for frequency and voltage regulation of microgrid has been proposed in this work. Accordingly, this paper develops a DOBC based frequency and voltage regulation strategy for solar PV-DG microgrid during islanding condition without using energy storage systems which account for a major component of the microgrid cost. Accordingly the proposed technique is well suited small and remote microgrids with limited energy generation and storage resources. Furthermore the control technique ensures efficient operation of microgrids with critical loads such as hospital and schools.\\
Major contributions of this paper are summarized as: \\
(1) Integration of a fast and robust auxiliary control to the existing conventional controllers, which improves the system performance without additional computational burden. \\
(2) Reduction of the frequency and voltage oscillations using a feed-forward control strategy based on DOBC. \\
(3) Robustness verification of the proposed control strategy under practical operating conditions. Accordingly, the system is tested under worst case uncertainty conditions, communication delay and noisy load conditions have been carried out.\\
(4) Implementation and verification of proposed control strategy on real-time simulator OPAL-RT and on laboratory scale microgrid setup. \\
\indent Remainder of the paper is organized as follows. System description and modeling of the various energy sources is presented in Section II. Section III describes the detailed modelling of DOBC and its application in frequency-voltage regulation. Section IV illustrates the results and discussions. Finally, major conclusions of the research are drawn in Section V. 
\section{System Description and Modeling}
\begin{figure}[h]\centering
\includegraphics[width=4.0in]{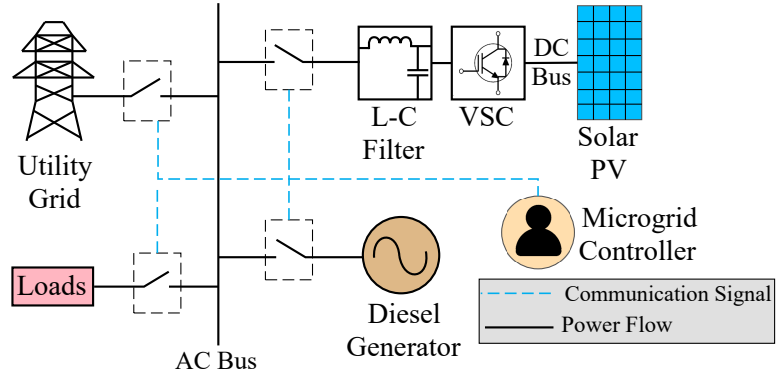}
\caption{Detailed description of a solar PV-Diesel microgrid.}
\label{microgrid}
\end{figure}

A schematic representation of a typical grid-connected solar PV-DG microgrid is illustrated in Fig. \ref{microgrid}. As can be seen from the fig., the microgrid consists of  generation sources including a solar PV generator and a DG which supply the loads along with grid power. Solar PV is connected to PCC through a voltage source converter (VSC). The VSC extracts the maximum power produced by solar PV using maximum power point tracking (MPPT) algorithm and operates it in grid-following mode, by tracking the frequency and voltage angle using phase-lock loop (PLL) at PCC. DG consists of a diesel engine coupled to synchronous machine which runs as a alternator. Also, it is assumed that all sources in microgrid are present at same geographical location. Control of the sources and loads in grid-connected/islanded modes of operation of the microgrid is managed by a microgrid central controller (MGCC). During the grid-connected mode, surplus/deficit power between solar PV \& load is managed by the utility grid, while the frequency and voltage of the microgrid are regulated by the utility grid, keeping the DG in disconnected mode. In islanding mode the load power requirement is managed by the solar PV and DG. This mode disconnects the microgrid from utility grid, while the frequency and voltage of the microgrid during the islanding mode are regulated by the synchronous machine of DG. 
The presence of line inductors in the LC filter with VSC significantly increases the X/R ratio of the microgrid. Consequently, variations in frequency does not directly impact the microgrid voltage, and vice-versa \cite{RAHMAN2016488}.
The dynamic modeling of various components of microgrid are described in the following subsections.
\subsection{Solar Photovoltaic (PV) Generator}
The output power of a solar photovoltaic generator is determined by the quantity of solar irradiation falling on the panel and its cell temperature. The relation of power generated by the solar PV generator with solar irradiance is expressed as \cite{8649677}.
\begin{equation}
P_{PV}(t)=S_I(t).\eta_a.A\\
\end{equation}
\begin{equation}
\eta_a(t)=\eta_r\eta_i[1-N_{Temp}(T_{PV}(t)-T_{ref})]    
\end{equation}
where $P_{PV}$ is the solar PV power output, $S_I(t)$ is the amount of solar incident irradiance at time \textit{t}. \textit{A} is the panel surface area and $\eta_a$ is the efficiency of solar PV array. $\eta_r$ is the module reference efficiency, $\eta_i$ is the VSC efficiency. $N_{Temp}$ is the PV panel temperature coefficient. $T_{PV}(t)$ is PV cell temperature at time \textit{t} and $T_{ref}$ is the reference temperature under standard testing conditions (STC). 
The PV cell temperature at any instant of time is calculated using the following equation:
\begin{equation}
T_{PV}(t)=T_a(t)+\frac{(NOCT-20)}{800}S_I(t)\
\end{equation}
where $T_a$ is the ambient temperature of air ($^\circ$C) and \textit{NOCT} is the nominal operating cell temperature.
At constant value of PV cell temperature $T_{PV}(t)$, the output of solar PV generator is found to be linearly varied with $S_I(t)$ \cite{6565075}. 
In this work it assumed that the temperature of solar PV cell remains constant and the output of solar PV generator is linearly varied by varying the solar irradiance level $S_I(t)$.
\subsection{Diesel Generator}
DG is widely used as backup energy source in renewable energy based microgrids to provide uninterrupted and secure power supply during islanding conditions of microgrids. DG is a essentially a synchronous generator which is coupled to diesel engine as the prime mover. The engine consumes diesel fuel as its source to drive the synchronous machine. In general during islanding conditions, DG regulates the voltage of the microgrid within specified limits using automatic voltage regulator (AVR). Further, the DG also regulates the frequency of the microgrid by maintaining the rotational speed of DG at synchronous speed \textit{Ns}, through load frequency control (LFC). 
\vspace{-0.3cm}  
\subsection{Microgrid System Configuration}
\begin{figure}[h]\centering
\includegraphics[width=4in]{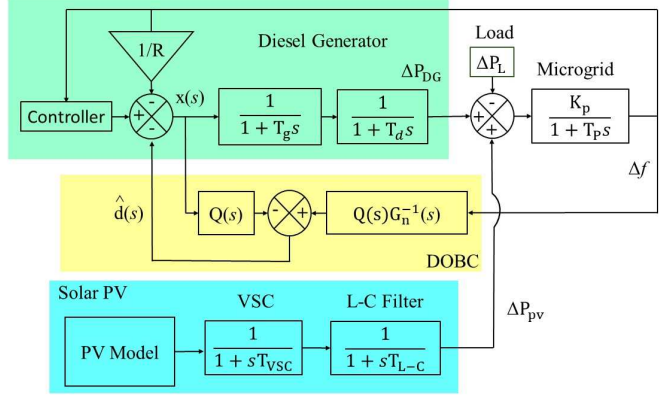}
\caption{Schematic diagram of dynamic model for frequency control during islanding.}
\label{LFC}
\end{figure}

Overall model of the islanded microgrid system for frequency control is shown in Fig. \ref{LFC} where, $\Delta P_{DG}$ is change in diesel generator power, \textit{R} is the droop coefficient, $T_{g}$ is turbine governor time constant, $T_{d}$ is diesel generator time constant. $\Delta P_{PV}$ is change in solar PV power, $T_{VSC}$ is VSC time constant, $T_{L-C}$ is L-C filter time constant, $\Delta P_{L}$ is change in load power, \textit{$\Delta$f} is frequency deviation, $K_p$ is power system gain and $T_p$ is power system time constant. The frequency deviation in power system is obtained by power system dynamics model and its parameters are calculated by the following equation\cite{osti_5599996};  
\begin{equation}
 K_p \triangleq\frac{1}{D} \;\;\;\;\;  Hz/pu \hspace{0.15cm}MW
\end{equation}
\begin{equation}
 T_p \triangleq \frac{2HD}{f^\circ}\;\;\;\; s
\end{equation}
where, \textit{D} is damping factor in \textit{Hz/pu MW}, \textit{$f^\circ$} is nominal frequency in \textit{Hz} and  \textit{H} is inertia constant in seconds, which is the ability of a system to resist change in frequency during any contingency. As evident from Fig. \ref{LFC}, any mismatch in the generated power and demand leads to frequency deviation in the system, which majorly depends on the system inertia (\textit{H}) and damping factor (\textit{D}) \cite{8676274}. 

\begin{figure}[h]\centering
\includegraphics[width=4in]{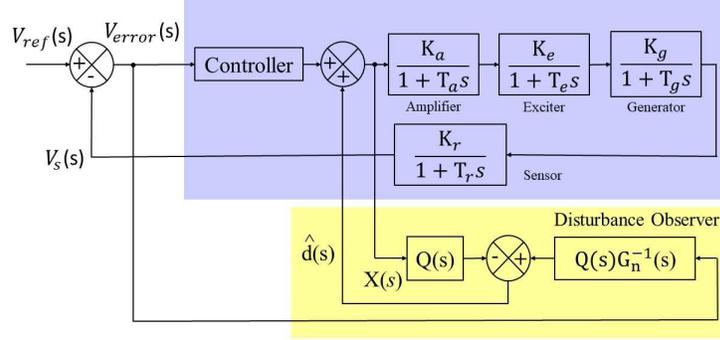}
\caption{Schematic diagram of dynamic model for voltage control during islanded microgrid}
\label{AVR}
\end{figure}

Similarly, the overall model of islanded microgrid system for voltage control is shown in Fig. \ref{AVR}, where \textit{$K_a$} is amplifier gain constant, $T_a$ is amplifier time constant, \textit{$K_e$} is exciter gain constant, \textit{$T_e$} is exciter time constant, $K_g$ is generator gain constant, $T_g$ is generator time constant, $K_r$ is sensor gain constant, $T_r$ is sensor time constant, $V_{ref}$ is reference terminal voltage in p.u., $V_{S}$ is sensed terminal voltage in p.u. $V_{error}$ is difference in reference terminal voltage $(V_{ref})$ and sensed terminal voltage.  
\section{Disturbance Observer Based Controller}
\subsection{Methodological Theory of DOBC Method}
\begin{figure}[h]\centering
\includegraphics[width=2.5in]{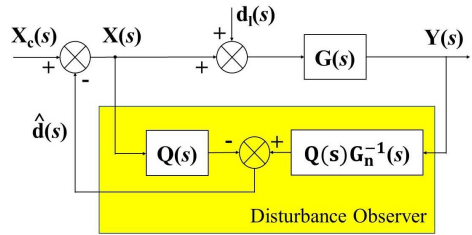}
\caption{Frequency domain disturbance observer architecture.}
\label{DOBC1}
\end{figure}

A linear disturbance estimator in frequency domain form has been considered as shown in Fig. \ref{DOBC1}.\\
\indent The DOBC architecture in frequency domain form is depicted as follows \cite{10.5555/2636749};
\begin{equation}
Y(s)=G(S)[X(s)+d_l(s)]    
\end{equation}
where $X(s)$ is the control input, $X_c(s)$is the controller output, $Y(s)$ is the controlled output, $G(s)$ is transfer function of plant model, $d_l(s)$ is the lumped disturbance, Q(s) is a filter transfer function and d\^(s) is the estimated equivalent disturbance. Since the disturbance observer can estimate both internal and external disturbances occurred due to model uncertainties, therefore, the DOBC architecture is modified to an equivalent diagram which consists of lumped disturbances as shown in Fig. \ref{DOBC1}. The lumped disturbance can be estimated as follows.
\begin{equation}
d_l(s)=G_n^{-1}(s)G(s)d(s)+[G_n^{-1}(s)G(s)-1]X(s)
\end{equation}
where $d(s)$ is the disturbance, $G^{-1}_n$ is the inverse of nominal plant transfer function. During the absence of any uncertainty in the system, the inverse of nominal plant transfer function ($G^{-1}_n(s)$) is same as the inverse of actual plant model ($G^{-1}(s)$). Estimated equivalent disturbance is calculated as follows.
\begin{equation}
\hat{d}(s)=Q(s)G_n^{-1}(s)Y(s)-Q(s)X(s)\\
\end{equation}
\begin{equation}
          =Q(s)G_n^{-1}(s)G_n(s)(X(s)+d_l(s))-Q(s)X(s)\\
\end{equation}
\begin{equation}
          =Q(s)d_l(s)
\end{equation}
Further, lumped disturbance estimation error $e_d(s)$ is calculated as difference in estimated equivalent disturbance ($\hat{d}(s)$) and lumped disturbance ($d_l(s)$). The value of lumped disturbance estimation error must tend to zero as the time tends to infinity, if the filter Q(s) is designed as a low-pass filter. The lumped disturbance estimation error ${e_d(s)}$ is calculated as follows.  
\begin{equation}
e_d(s)=\hat{d}(s)-d_l(s)=[Q(s)-1]d_l(s)\\
\end{equation}
\begin{equation}
=[Q(s)-1]d_l(s)
\end{equation}

\subsection{Application of Linear Disturbance Estimator in AVR.}
In order to provide voltage regulation support during islanding condition, the designing and implementation of DOBC is explained in this section. As shown in Fig. \ref{DOBC1}, it is assumed that the control input $\big(X(s)\big)$ is affected by disturbance $\big(d(s)\big)$ in the system, but for AVR application the control instruction is not directly affected by the disturbance $\big(d(s)\big)$.
The disturbance in the system is resultant of voltage mismatch, 
\begin{equation}
d(s)={V_{ref}(s)- V_s(s)}
\end{equation} 
The implementation of DOBC for AVR includes two more procedures which are as follows:
1) Identification of G(s).
2) Designing of Q(s). 

1) Identification of G(s):
The overall plant transfer function is given as,
\begin{equation}
G(s)=\Bigg[\frac{K_a}{1+sT_a}\Bigg]\Bigg[\frac{K_e}{1+sT_e}\Bigg]\Bigg[\frac{K_g}{1+sT_g}\Bigg]\Bigg[\frac{K_r}{1+sT_r}\Bigg]
\end{equation}
Substituting the values of the above parameters, overall plant model is calculated as,
\begin{equation}
G(s)=\Bigg[\frac{10}{1+1.51s+0.555s^2+0.0454s^3+0.004s^4}\Bigg]
\end{equation}

\begin{figure}[h]\centering
\includegraphics[width=3.75in,height=2.0in]{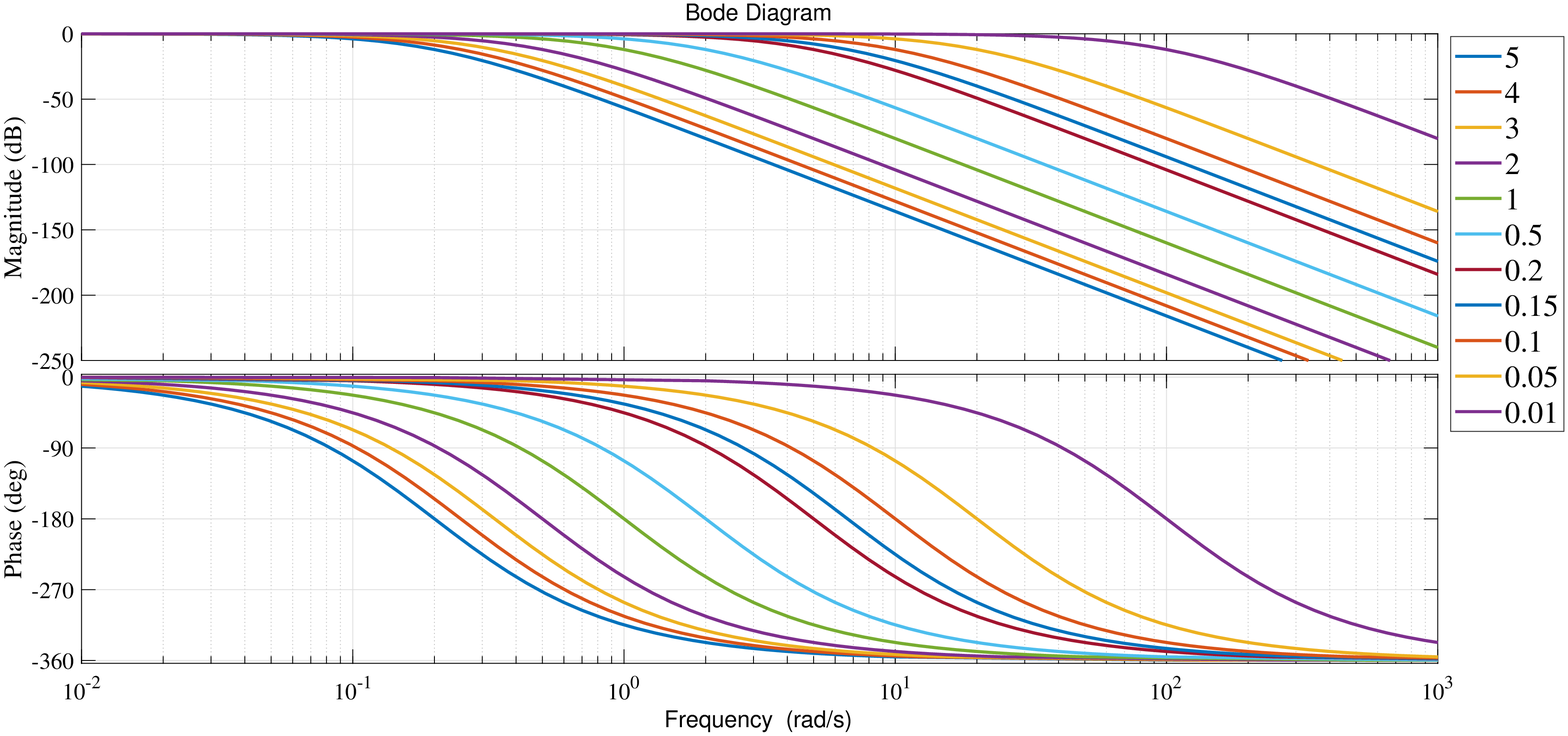}
\caption{Bode diagram of filter Q(s) under different values of $\lambda$.}
\label{bodeplotofavrfilter}
\end{figure}

2) Designing of Q(s):
The efficiency of DOBC is majorly dependent on the designing of Q(s). It must be designed as a low-pass filter such that it eliminates the high frequency noise generated by the sensor. The order of Q(s) must not be less than the difference in order of denominator and numerator of plant model (G(s)). The transfer function of Q(s) selected in this case is represented as a fourth order transfer function given by;
\begin{equation}
Q(s)=\frac{1}{(\lambda s+1)^4}
\end{equation}
The filter accuracy in estimating the disturbance depends upon the selection of filter parameter lambda ($\lambda$). The bode plot of filter Q(s) under different values of $\lambda$ is depicted in Fig. \ref{bodeplotofavrfilter}. It is evident from the figure that the frequencies higher than $10^{-1} rad/s$ are attenuated. The presence of phase lag degrades the performance of the filter, therefore the filter must have zero lag. The values of gain, phase and cut-off frequency at different values of $\lambda$ are shown in Table \ref{table1}. It can observed that by reducing the value of $\lambda$ the gain tends to unity, phase lag becomes nearly zero and cut-off frequency increases. Appropriately the value of $\lambda$ has been considered as 0.01 which satisfies the conditions of unity gain and zero phase lag. Similar procedure is carried out for the application of linear disturbance estimator in LFC. Due to space constraints the detailed analysis is not shown in this paper.
\begin{table}
\centering\caption{Filter performance at different values of $\lambda$.}\label{table1}
\footnotesize
	\begin{tabular}{c c c c}
		\hline 
		\hline 		
		\rule{0pt}{2ex} \textbf{Lambda} & \textbf{Gain(dB)} & \textbf{Phase(degree)} & \textbf{Cut-off frequency} \\
		\rule{0pt}{2ex} \textbf{$\lambda$} & \textbf{$@10^{-1}rad/sec$} & \textbf{$@10^{-1}rad/sec$} & \textbf{rad/sec} \\
		\hline
		\rule{0pt}{2ex} 5 & -4.37 & -113 & 0.091 \\
		\rule{0pt}{2ex} 4 & -2.47 & -85.4 & 0.097 \\
		\rule{0pt}{2ex} 3 & -1.84 & -74 & 0.111 \\
		\rule{0pt}{2ex} 2 & -0.743 & -47.2 & 0.122 \\
		\rule{0pt}{2ex} 1 & -0.216 & -25.5 & 0.179 \\
		\rule{0pt}{2ex} 0.5 & -0.045 & -11.7 & 0.669 \\
		\rule{0pt}{2ex} 0.2 & -0.006 & -4.58 & 1.05 \\
		\rule{0pt}{2ex} 0.15 & -0.006  & -4.58 & 1.39 \\
		\rule{0pt}{2ex} 0.1 & $-6.95*10^{-5}$ & -0.458 & 2.45 \\
		\rule{0pt}{2ex} 0.05 & $-6.95*10^{-5}$  & -0.458 & 5.72  \\
		\rule{0pt}{2ex} 0.01 & $-6.95*10^{-5}$ & -0.458 & 28.6 \\
		\hline
	\end{tabular}
\end{table}

\section{Results and Discussion}
\subsection{Real-time Simulation}
\begin{figure}[h]\centering
\includegraphics[width=3in]{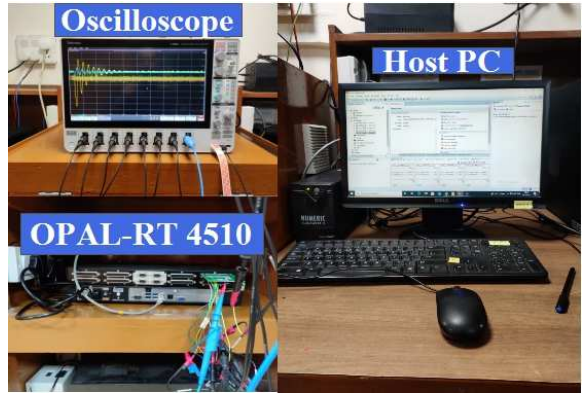}
\caption{Laboratory setup demonstrating the real-time validation of proposed controller using OPAL-RT. }
\label{OPALRT}
\end{figure}

The proposed control strategy for the microgrid shown in Fig. \ref{microgrid} is developed on MATLAB/Simulink platform and results are tested using OPAL-RT real-time simulator. The real-time laboratory setup is shown in Fig. \ref{OPALRT}, which consists of a host-computer(PC) for interfacing MATLAB/Simulink with OPAL-RT 4510 and a eight channel mixed-channel oscilloscope for acquiring the waveforms. The control strategy is run on real-time platform with a fixed time step of 10 $\mu$s. The typical values of the parameters used in modelling for frequency control as shown in Fig. \ref{LFC} are suggested in \cite{5677458,8398456} and similarly for voltage control as shown in Fig. \ref{AVR} are suggested in \cite{SOLIMAN2021107216}. The typical values of system parameters are provided in Appendix. The steady state and transient performance of the controllers are compared using performance indices such as integral of the square of the error (ISE), integral time-multiplied of the square of the error (ITSE), integral of absolute value of error (IAE), integral of the time-multiplied absolute value of error (ITAE), maximum overshoot (MO) and settling time.

\subsection{Performance analysis of AVR}
Performance of the proposed control strategy i.e. integration of DOBC with Proportional-Integral-Derivative (PID) controller for terminal voltage regulation is evaluated in this section. The performance of the proposed DOBC integrated controller is compared with the different PID controllers whose parameters have been obtained in literature \cite{SOLIMAN2021107216}.

\begin{figure}[h]\centering
\includegraphics[width=3in]{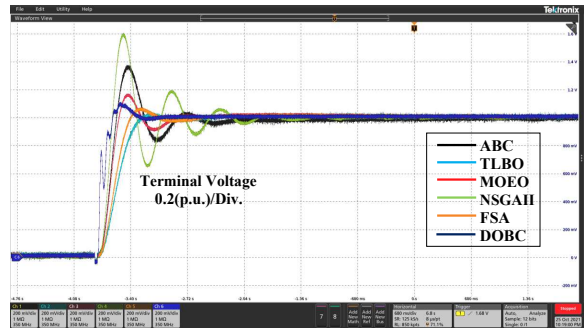}
\caption{Comparison of terminal voltage response. }
\label{caseaavr}
\end{figure}

\begin{figure}[h]\centering
\includegraphics[width=3in]{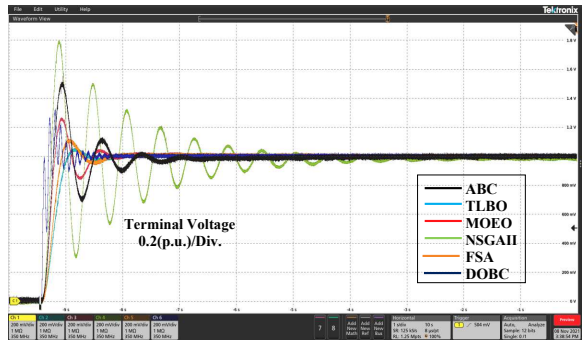}
\caption{Comparison of terminal voltage response under communication delay.}
\label{casebavr}
\end{figure}

Case A: Unit step change in $V_{ref}$.\\
A unit step change of 1.0 p.u. in reference voltage of DG is provided and the comparison of terminal voltage response is shown in Fig. \ref{caseaavr}. The settling time within $5\%$ of dead-band is found to be minimum for the proposed control scheme, and comparable values are obtained for other PI controller comparison of the steady state and transient response of terminal voltage is depicted in Table \ref{comparisonAVR}. 
\\
Case B: Unit step change in $V_{ref}$ under  communication delay.\\
In a practical microgrid, the information of electrical parameters are provided through sensors and the information is communicated to microgrid central controller through a communication channel. The communication channel can be a shared medium and may face communication delays and uncertainties. Therefore, it is essential to verify robustness of the proposed control in the presence of communication delays. Accordingly a scenario with communication delay of 0.02s has been considered in this work. Comparison of the performance of terminal voltage response for step change of 1.0 p.u. under communication delay is depicted in Fig. \ref{casebavr}. 
The settling time within $5\%$ of dead-band is found to be minimum for the proposed control scheme and comparable values are obtained for the other PI controllers. A comparison of the steady state and transient responses of terminal voltage is depicted in Table \ref{comparisonAVR}. 
\begin{table}[h]
\caption{Comparison of steady state and transient response of terminal voltage.}
\label{comparisonAVR}
\centering \footnotesize
\begin{tabular}{|c| c c c c c c|}
     \hline
        \hline 
        \multicolumn{7}{|c|}{\textit{Case A}}\\
        \hline
        \hline
        \rule{0pt}{2ex}  & $ISE$ & $ITSE$ & $IAE$ & $ITAE$ & $MO$ & Settling Time \\
         & & & & & (p.u.) & (Seconds)\\ 
        \hline
        \rule{0pt}{2ex} $ABC$ & $0.145$ & $0.024$ & $0.346$ & $0.298$ & $1.354$ & $1.402$ \\
        \hline
        \rule{0pt}{2ex} $TLBO$ & $0.183$ & $0.021$ & $0.273$ & $0.079$ & $1.006$ & $0.469$ \\
        \hline 
        \rule{0pt}{2ex} $MOEO$ & $0.127$ & $0.011$ & $0.228$ & $0.098$ & $1.153$ & $0.818$ \\
        \hline
        \rule{0pt}{2ex} $NSGAII$ & $0.171$ & $0.041$ & $0.411$ & $0.323$ & $1.571$ & $1.81$ \\
        \hline 
        \rule{0pt}{2ex} $FSA$ & $0.161$ & $0.016$ & $0.243$ & $0.065$ & $1.049$ & $0.523$ \\
        \hline
        \rule{0pt}{2ex} $DOBC$ & $0.039$ & $0.001$ & $0.089$ & $0.016$ & $1.086$ & $0.371$ \\
        \hline
        \multicolumn{7}{|c|}{\textit{Case B}}\\
        \hline
        \hline
        \rule{0pt}{2ex}  & $ISE$ & $ITSE$ & $IAE$ & $ITAE$ & $MO$ & Settling Time \\
         & & & & & (p.u.) & (Seconds)\\ 
        \hline
        \rule{0pt}{2ex} $ABC$ & $0.199$ & $0.049$ & $0.485$ & $0.614$ & $1.497$ & $1.569$ \\
        \hline
        \rule{0pt}{2ex} $TLBO$ & $0.183$ & $0.021$ & $0.273$ & $0.081$ & $1.006$ & $0.9524$ \\
        \hline 
        \rule{0pt}{2ex} $MOEO$ & $0.154$ & $0.018$ & $0.285$ & $0.131$ & $1.257$ & $0.845$ \\
        \hline
        \rule{0pt}{2ex} $NSGAII$ & $0.397$ & $0.261$ & $0.949$ & $1.404$ & $1.8$ & $4.255$ \\
        \hline 
        \rule{0pt}{2ex} $FSA$ & $0.179$ & $0.019$ & $0.272$ & $0.080$ & $1.104$ & $0.955$ \\
        \hline
        \rule{0pt}{2ex} $DOBC$ & $0.055$ & $0.002$ & $0.108$ & $0.020$ & $1.225$ & $0.369$ \\
        \hline
        \hline
    \end{tabular}
\end{table}
\subsection{Performance analysis of LFC.}
Performance of the proposed control strategy i.e. integration of DOBC with Integral controller for LFC is evaluated in this section.\\
Case A: Performance evaluation under different robustness tests.\\
The robustness of the proposed control is evaluated  by testing under maximum uncertainty occurrence in generation and load, obtained as under the worst-case uncertainties. Worst-case scenario are identified considering polyhedral uncertainty sets for PV output and load under various test conditions. The polyhedral uncertainty sets for PV output and load are represented as follows
\begin{equation}
 U_{PV}= \Big\{{{P}^{PV,t}}\in{\mathbb{R}^{n_{PV}}} :{P^{PV,t}_{min}} \leq P^{PV,t} \leq P^{PV,t}_{max}, \forall\;{t}\;{\zeta^{l}_{PV}} \leq \frac{{\sum_{t\in T}P^{PV,t}}}{{\sum_{t\in T}P^{PV,t}_f}} \leq \zeta^{u}_{PV} \Big\}
\end{equation}
\begin{equation}
 U_{L}= \Big\{{{P}^{L,t}}\in{{\mathbb{R}^{n_{L}}} :{P^{L,t}_{min}}} \leq P^{L,t} \leq P^{L,t}_{max}, \forall\;{t}\;{\zeta^{l}_{L}} \leq \frac{{\sum_{t\in T}P^{L,t}}}{{\sum_{t\in T}{P^{L,t}_f}}} \leq \zeta^{u}_{L} \Big\}
\end{equation}

\begin{figure}\centering
\includegraphics[width=3in]{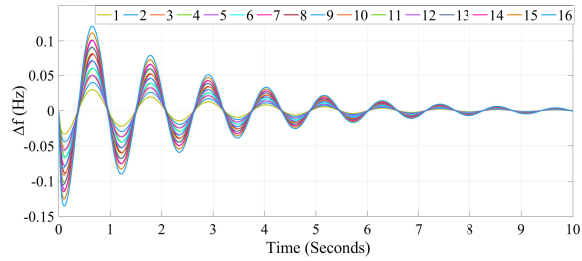}
\caption{Frequency response under test conditions.}
\label{caseafrequency}
\end{figure}

Here $\zeta ^l_{PV}$ \& $\zeta ^u_{PV}$ are the lower and upper uncertainty budgets of solar PV in p.u., respectively. $\Delta P_{PV}$ refers to the step change in solar PV power in p.u., considering base value of DG. Similarly $\zeta ^l_{L}$ \& $\zeta ^u_{L}$ are the lower and upper uncertainty budgets of load in p.u., respectively.  $\Delta P_{L}$ denotes a step change in load power in p.u. considering base value of DG. $\Delta f_{max}$ is the maximum frequency deviation in Hz.\\
The different test conditions considered for evaluation of the proposed technique are depicted in Table \ref{testconditions}. The table also shows the comparison of maximum frequency deviations under different test conditions considering different uncertainty budgets. The frequency response under the different test conditions is illustrated in Fig. \ref{caseafrequency}. it is evident from the results summarized in the table that $\Delta f$ is found to be maximum for test no. 16. Thus, the step change conditions in solar PV output and load for this test are observed to be the worst case uncertainty scenario for further simulations.   
\begin{table}[h]
\caption{Frequency deviation under different test conditions.}
\label{testconditions}
\centering\footnotesize
    \begin{tabular}{c c c c c c c c}
        \hline
        \hline
         $Test No.$ & $\zeta ^l_{PV}$ & $\zeta ^u_{PV}$ & $\Delta P_{PV}$ & $\zeta ^l_{L}$ & $\zeta ^u_{L}$ & $\Delta P_{L}$ & $\Delta f_{max}$ \\
         & $(p.u.)$ & $(p.u.)$ & $(p.u.)$ & $(p.u.)$ & $(p.u.)$ & $(p.u.)$ & $(Hz)$ \\
        \hline
        \rule{0pt}{2ex} $1$ & $0.9$ & $1.1$ & $0.016$ & $0.9$ & $1.1$ & $0.033$ & $0.033$ \\
        \rule{0pt}{2ex} $2$ & $0.8$ & $1.2$ & $0.033$ & $0.9$ & $1.1$ & $0.033$ & $0.044$ \\
        \rule{0pt}{2ex} $3$ & $0.7$ & $1.3$ & $0.05$ & $0.9$ & $1.1$ & $0.033$ & $0.055$  \\
        \rule{0pt}{2ex} $4$ & $0.6$ & $1.4$ & $0.066$ & $0.9$ & $1.1$ & $0.033$ & $0.066$ \\
        \rule{0pt}{2ex} $5$ & $0.9$ & $1.1$ & $0.016$ & $0.8$ & $1.2$ & $0.066$ & $0.056$ \\
        \rule{0pt}{2ex} $6$ & $0.8$ & $1.2$ & $0.033$ & $0.8$ & $1.2$ & $0.066$ & $0.067$   \\
        \rule{0pt}{2ex} $7$ & $0.7$ & $1.3$ & $0.05$ & $0.8$ & $1.2$ & $0.066$ & $0.078$ \\
        \rule{0pt}{2ex} $8$ & $0.6$ & $1.4$ & $0.066$ & $0.8$ & $1.2$ & $0.066$ & $0.086$ \\
        \rule{0pt}{2ex} $9$ & $0.9$ & $1.1$ & $0.016$ & $0.7$ & $1.3$ & $0.1$ & $0.081$ \\
        \rule{0pt}{2ex} $10$ & $0.8$ & $1.2$ & $0.033$ & $0.7$ & $1.3$ & $0.1$ & $0.091$ \\
        \rule{0pt}{2ex} $11$ & $0.7$ & $1.3$ & $0.05$ & $0.7$ & $1.3$ & $0.1$ & $0.102$ \\
        \rule{0pt}{2ex} $12$ & $0.6$ & $1.4$ & $0.066$ & $0.7$ & $1.3$ & $0.1$ & $0.112$ \\
        \rule{0pt}{2ex} $13$ & $0.9$ & $1.1$ & $0.016$ & $0.6$ & $1.4$ & $0.133$ & $0.105$ \\
        \rule{0pt}{2ex} $14$ & $0.8$ & $1.2$ & $0.033$ & $0.6$ & $1.4$ & $0.133$ & $0.115$ \\
        \rule{0pt}{2ex} $15$ & $0.7$ & $1.3$ & $0.05$ & $0.6$ & $1.4$ & $0.133$ & $0.125$ \\
        \rule{0pt}{2ex} $16$ & $0.6$ & $1.4$ & $0.066$ & $0.6$ & $1.4$ & $0.133$ & $0.135$ \\
        \hline
        \hline
    \end{tabular}
\end{table}

\begin{figure}\centering
\includegraphics[width=3in]{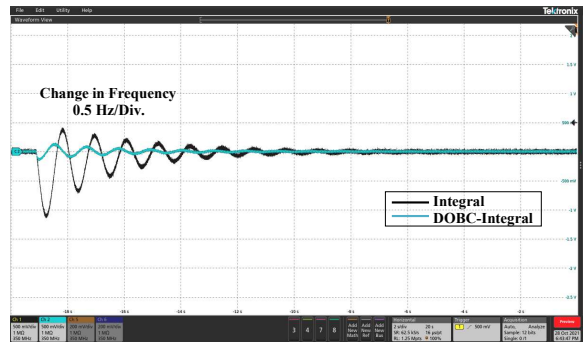}
\caption{Frequency response for a worst case uncertainty scenario.}
\label{casebfrequency}
\end{figure}

\begin{figure}\centering
\includegraphics[width=3in]{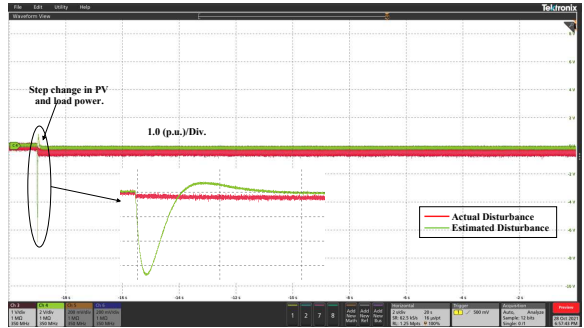}
\caption{Actual and estimated disturbance response for a worst case uncertainty scenario.}
\label{casebdisturbance}
\end{figure}

Case B: Frequency response under worst case uncertainty scenario.\\
The worst case scenario as identified in case A, with a step change of 0.066 p.u. in solar PV output and 0.133 p.u. in load power has been considered to test the proposed control strategy. The frequency response of the proposed control strategy under step change is compared with the conventional integral controller and the results are illustrated in Fig. \ref{casebfrequency}. The actual and estimated disturbance response of the proposed strategy are shown in Fig. \ref{casebdisturbance}. It can observed from the Fig. \ref{casebdisturbance}, that the proposed strategy can efficiently track the disturbances within a short period of time.
Further, a comparison of the steady state and transient responses of LFC are depicted in Table \ref{comparisonLFC}. It is evident that the performance indices are found to be minimum for the proposed control as compared with Integral control.  

\begin{figure}\centering
\includegraphics[width=3in]{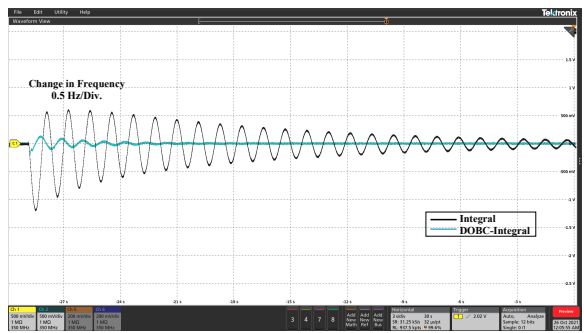}
\caption{Frequency response under communication delay.}
\label{casecfrequency}
\end{figure}

\begin{figure}\centering
\includegraphics[width=3in]{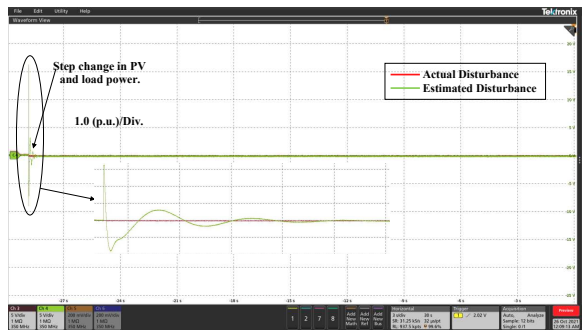}
\caption{Actual and estimated disturbance response under communication delay.}
\label{casecdisturbance}
\end{figure}

Case C: Frequency response under communication delay.\\
In order to further verify the robustness of the proposed control strategy in practical operating conditions, the system has been tested under the presence of communication delay. A communication delay of 0.02 seconds has been considered for a step change of 0.066 p.u. in solar PV output and 0.133 p.u. in load power. The frequency response under the communication delay is shown in Fig. \ref{casecfrequency}. The actual and estimated disturbance responses under step change are shown in Fig. \ref{casecdisturbance}. Further, a comparison of the steady state and transient responses of LFC are depicted in Table \ref{comparisonLFC}. Results reveal that the performance indices are found to be minimum for the proposed control, as compared with Integral control. 
\begin{figure}\centering
\includegraphics[width=3in]{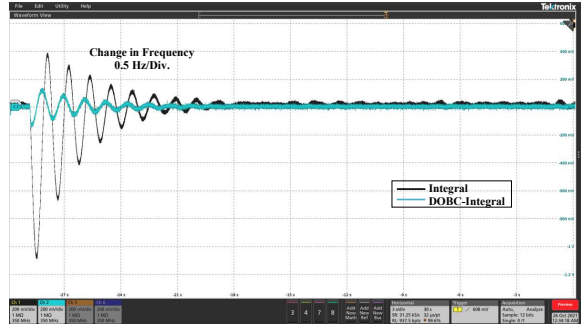}
\caption{Frequency response considering white noise in load.}
\label{casedfrequency}
\end{figure}
\begin{figure}\centering
\includegraphics[width=3in]{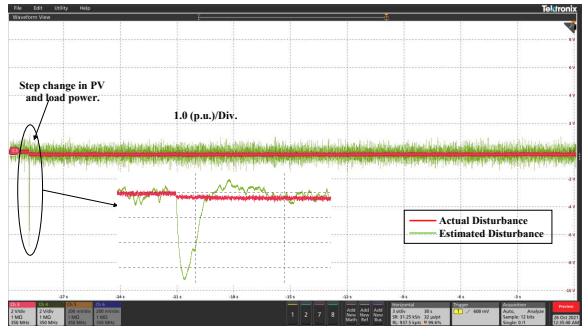}
\caption{Actual and estimated disturbance response considering white noise in load.}
\label{caseddisturbance}
\end{figure}
\begin{table}
\caption{Comparison of steady state and transient response of LFC.}
\label{comparisonLFC}\centering\footnotesize
    \begin{tabular}{c c c c c c c }
        \hline
        \hline 
        \multicolumn{7}{c}{Case B}\\
        \hline
        \hline
        \rule{0pt}{2ex}  & $ISE$ & $ITSE$ & $IAE$ & $ITAE$ & $\Delta f_{max}$ & Settling Time \\
         & & & & & (Hz) & (Seconds)\\ 
        \hline
        \rule{0pt}{2ex} $Integral$ & $0.738$ & $1.575$ & $1.619$ & $5.566$ & $1.108$ & $8.271$ \\
        \rule{0pt}{2ex} $DOBC$ & $0.0147$ & $0.0341$ & $0.25$ & $0.891$ & $0.135$ & $3.97$ \\
        \hline 
        \multicolumn{7}{c}{Case C}\\
        \hline
        \hline
        \rule{0pt}{2ex} $Integral$ & $2.613$ & $19.47$ & $6.704$ & $74.06$ & $1.19$ & $29.41$ \\
        \rule{0pt}{2ex} $DOBC$ & $0.016$ & $0.036$ & $0.257$ & $0.930$ & $0.189$ & $3.96$ \\
        \hline 
        \multicolumn{7}{c}{Case D}\\
        \hline
        \hline
        \rule{0pt}{2ex} $Integral$ & $0.739$ & $1.584$ & $1.623$ & $5.534$ & $1.105$ & $8.26$  \\
        \rule{0pt}{2ex} $DOBC$ & $0.014$ & $0.034$ & $0.252$ & $0.902$ & $0.134$ & $3.96$ \\
        \hline
        \hline
    \end{tabular}
\end{table}

Case D: Frequency response considering white noise in load.\\
Frequent switching operation of loads in the power system causes white noise in the loads. Considering the presence of white noise in the load, a step change of 0.066 p.u. in solar PV output and 0.133 p.u. in load power are considered as a test case. The frequency response under noisy load conditions is shown in Fig. \ref{casedfrequency}. The actual and estimated disturbance responses under step change are shown in Fig. \ref{caseddisturbance}. Further, a comparison of the steady state and transient responses of LFC are depicted in Table \ref{comparisonLFC}. It is observed that the performance indices are found to be minimum for the proposed control as compared with Integral control. 
\subsection{Hardware experimental verification}
\begin{figure}\centering
\includegraphics[width=\linewidth]{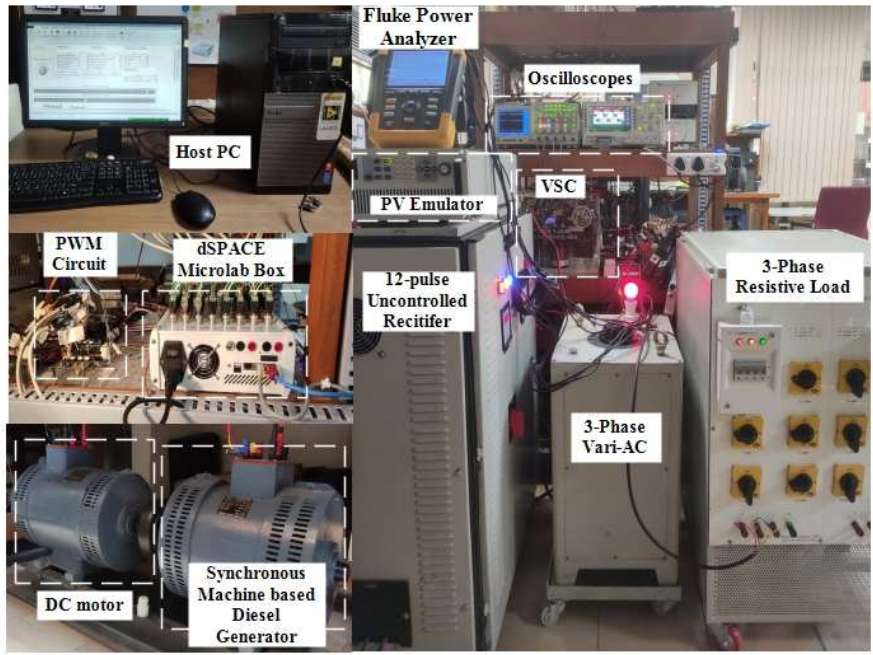}
\caption{Laboratory scale hardware setup.}
\label{setup}
\end{figure}

In order to verify the effectiveness of the proposed control strategy, a laboratory experimental setup has been developed as shown in Fig. \ref{setup}. The microgrid configuration is shown Fig. \ref{setup}, which consists of DC machine coupled synchronous generator to mimic the behaviour of DG. The field voltage of DC machine is regulated at fixed voltage, whereas the armature voltage is regulated through DC-DC buck converter connected to 12-pulse uncontrolled rectifier. The armature voltage is regulated based on feedback of frequency calculated through voltage measurements using phase-lock-loop (PLL). A delta-star isolation transformer is connected at the output of DG and three-phase resistive load are connected at PCC as shown in Fig. \ref{microgrid}. A PV emulator is connected to DC-DC boost converter, which tracks the maximum power point (MPP) of solar PV using incremental conductance method (INC). The solar PV power is feed to a 6-pulse 3-phase voltage source converter (VSC), which converters DC power generated by solar PV to PWM voltages. The interfacing inductors $(L_{abc})$ and ripple filter $(C_{abc})$ are connected at the output of VSC to filter the PWM voltages. The gate pulses and analog input task is done through dSPACE microlab box at sampling time of 60 $\mu S$. The VSC control algorithm regulates the DC link voltage and injects the power to load at unity power factor using hysteresis control. The typical values of system parameters for hardware experimental verification are provided in Appendix.
\begin{figure}\centering
\includegraphics[width=\linewidth]{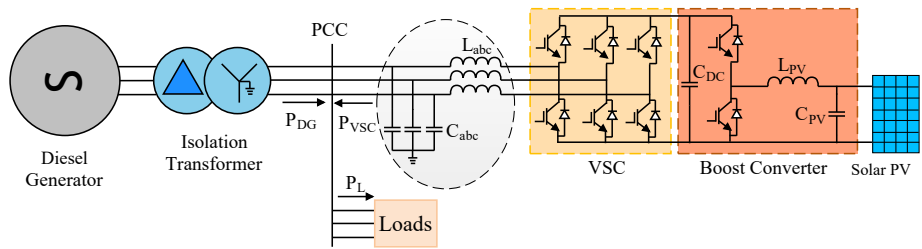}
\caption{Microgrid Configuration.}
\label{configuration}
\end{figure}

\subsection{Plant Model Identification and Stability Analysis}
\begin{figure}\centering
\includegraphics[width=\linewidth]{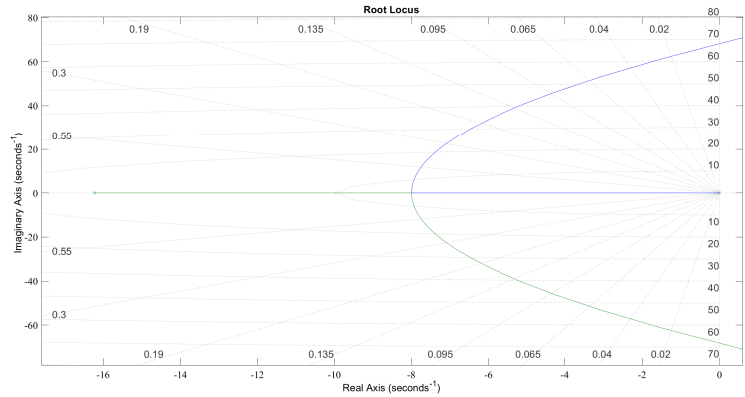}
\caption{Pole-Zero map of plant transfer function with feedback loop.}
\label{pole}
\end{figure}

The developed control strategy is dependent on system plant function and its experimental implementation requires  identification of system plant model. The identification of overall plant transfer function is performed by logging the duty cycle data which is sent to buck converter for armature control of DC machine with load frequency values. The logged data is imported to system identification tool provided in MATLAB. The overall plant transfer function of second order with estimated accuracy of 98.3$\%$ is given as;
\[
    G(s) = \frac{2.68e5}{s^2+303.4s+4661}
\]
A second order transfer function of filter (Q(s)) is considered in hardware verification, whose transfer function is given by:
\begin{equation}
Q(s)=\frac{1}{(\lambda s+1)^2}
\end{equation}
Appropriately, the value of $\lambda$ has been considered as 0.02, which satisfies the conditions of unity gain and zero phase lag. The integral value gain of controlled transfer function is taken as 0.01, identified using Routh-Hurwitz method. Also, the stability analysis of plant function is shown in Fig. \ref{pole}. The Figure shows the pole-zero map of the overall plant transfer feedback function, since the pole lies on the left side of s-plane, the stability of estimated control system is verified. 

\subsection{Experimental Performance Evaluation Under Worst Case Uncertainty Scenario.}
\begin{figure}\centering
\includegraphics[width=\linewidth]{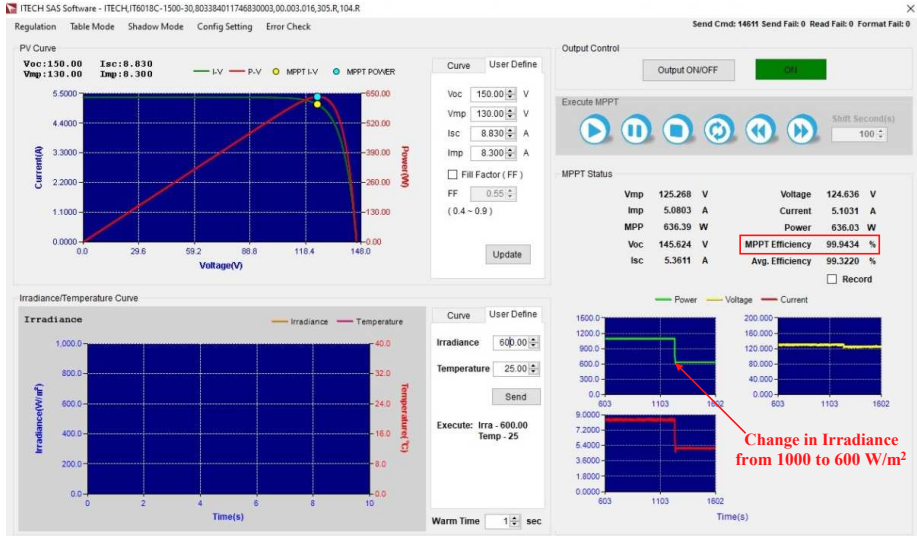}
\caption{Performance of INC-MPPT under dynamic irradiance conditions.}
\label{PV}
\end{figure}

\begin{figure}\centering
\includegraphics[width=0.7\linewidth]{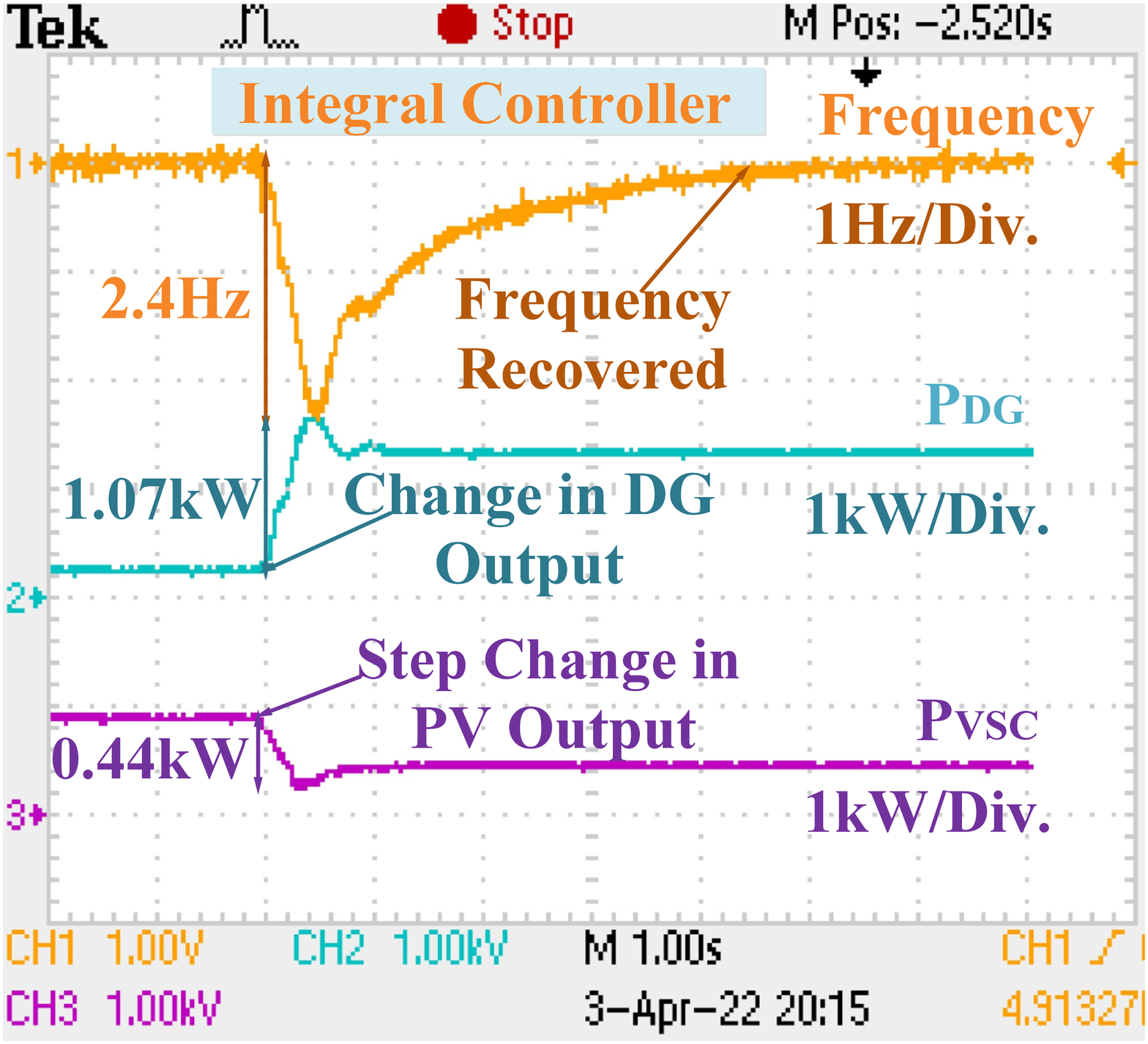}
\caption{Dynamic performance of integral control for LFC under step change in load and PV.}
\label{integral}
\end{figure}

\begin{figure}\centering
\includegraphics[width=0.7\linewidth]{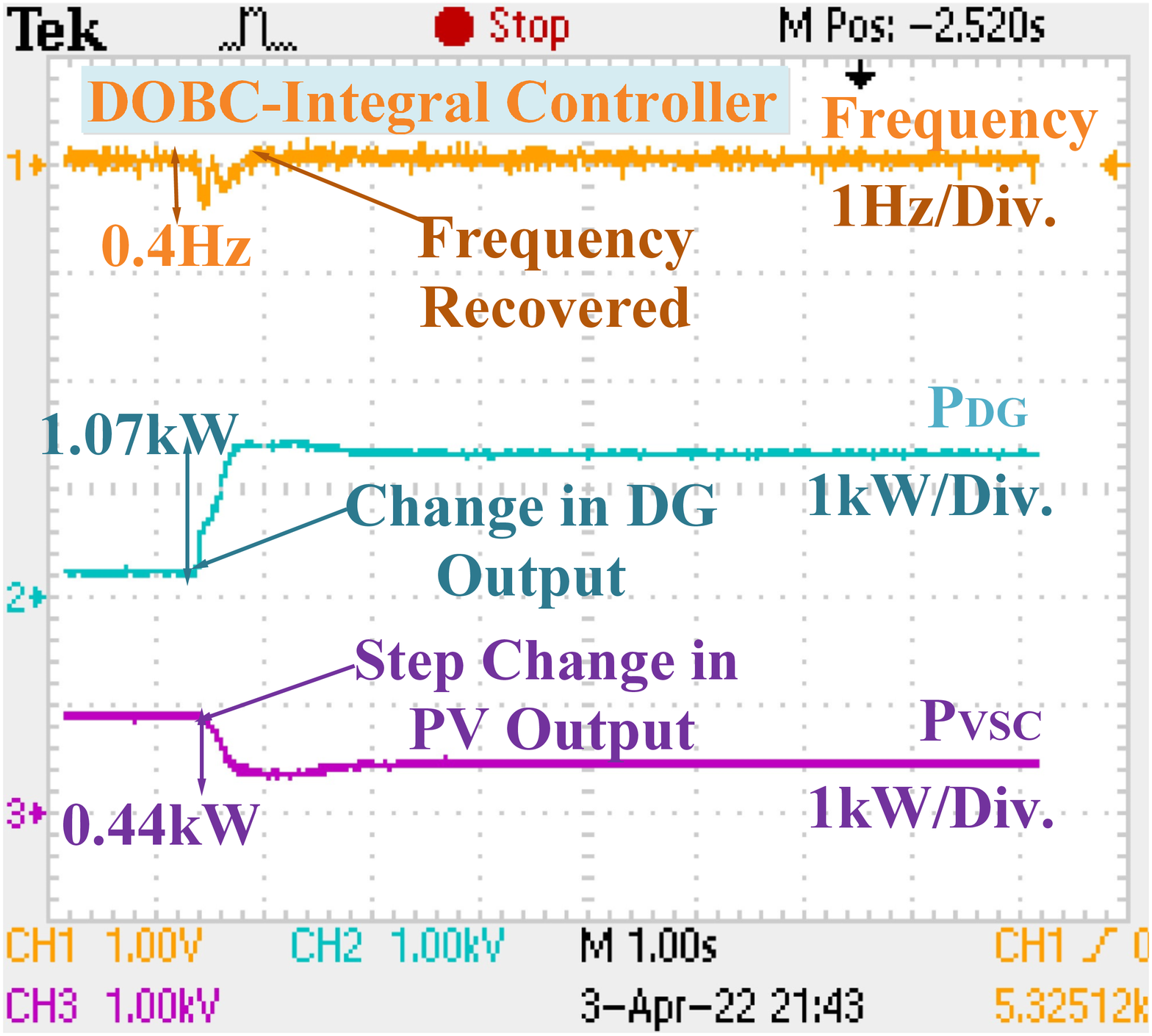}
\caption{Dynamic performance of DOBC-Integral control for LFC under step change in load and PV.}
\label{DOBC}
\end{figure}

\begin{figure}\centering
\includegraphics[width=0.7\linewidth]{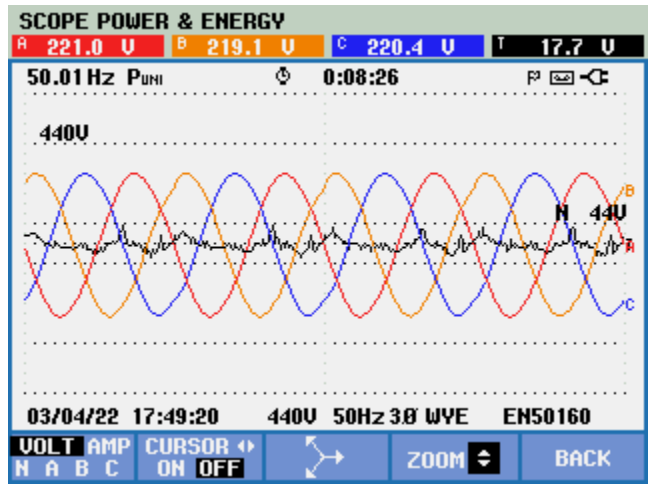}
\caption{Voltage waveform at PCC.}
\label{voltage}
\end{figure}

\begin{figure}\centering
\includegraphics[width=0.7\linewidth]{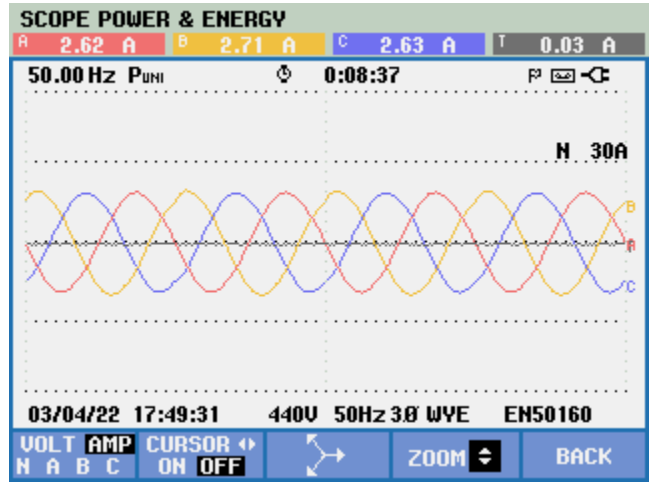}
\caption{Load current waveform after step change.}
\label{current}
\end{figure}

The performance of integrating DOBC with integral controller has been verified in this case. The VSC operates in grid forming mode and maintains the DC link voltage. The synchronous machine regulates the system frequency and voltage using LFC and AVR control. The integral controller is used in both LFC and AVR. The synchronous machine is nominally operated at 0.242 kW of active power and VSC generates an active power of 0.886 kW at solar irradiance of $1000 W/m^2$ to feed three-phase resistive load bank. To verify the effectiveness of proposed control strategy under worst case uncertainty scenario, a step change of 0.635 kW in load and a step change of 0.44 kW of VSC power has been observed due to change in solar irradiance from $1000 W/m^2$ to $600 W/m^2$ as shown in Fig. \ref{PV}. The performance of integral controller for LFC under step change in load and solar PV output is shown in Fig. \ref{integral}. A frequency change of 2.4 Hz is observed as shown in figure and it takes around 4.5 seconds for the frequency to recover by supplying active power through DG. Similar step change condition of solar PV and load is provided with  DOBC-Integral controller, and a frequency deviation of 0.4 Hz is observed from the Fig. \ref{DOBC} with a recovery time less than a second. As it can be observed from dynamic response of frequency that peak undershoot of frequency is reduced drastically during the step change when DOBC is integrated to Integral controller. Such reduction in frequency is due to fast and accurate estimation of disturbance using well designed DOBC filters. Fast and accurate detection of error relatively helps DG to supply active power mismatch without adding much computational burden as compared to slow integrating action of Integral controller. The load voltage waveform at PCC is shown in Fig. \ref{voltage} and similarly load current waveform after step load change is shown in Fig. \ref{current}.

\section{Conclusion}
The papers develops a frequency and voltage regulation scheme for a solar PV-DG microgrid under islanding conditions. The proposed control technique integrates the DOBC as a feed-forward controller to the frequency and voltage controller of DG. The microgrid is further integrated with a solar PV generator in parallel to the DG, which inherently introduces operational uncertainties and challenges to the microgrid. In view of this disturbances, the developed control scheme appropriately regulates the load frequency and voltage in presence of disturbances occurring due to the intermittent nature of source and variability in load. The performance of the developed control scheme has been validated through several robustness test simulation conditions using MATLAB/Simulink under varied uncertainty conditions. Performance of the proposed controller is further evaluated considering the impact of communication delays and white noise, representative of practical operating conditions, A comprehensive comparison of the performance indices of the proposed control with conventional controllers reveals that the proposed strategy provides better performance without an increased computational burden. Real-time performance of the proposed controller has been analysed through practical implementation using real-time simulator OPAL-RT. The real-time simulation results verifies the suitable performance of the proposed controller under practical operating conditions. Finally, the practical validation of proposed control is verified through laboratory scale microgrid setup, which verifies the frequency deviation reduction with integration of DOBC Controller to conventional controller without adding any additional sensor or computational burden. 
\section*{Appendix A: System Parameters for Real-time Simulation}
{\textit{Diesel Generator}}\\ Rated Power= 3kVA; Voltage= 415$V_{L-L}$; R= 2.4; $T_g$= 0.0728; $T_d$= 0.273; $f^o$= 50 H.z.; H= 1.0s; D= 0.0067.
$K_a$= 10; $T_a$= 0.1; $K_e$= 1; $T_e$= 0.4; $K_g$= 1; $T_g$= 1; $K_r$= 1; $T_r$= 0.01;\\
\indent {\textit{Solar PV}}\\ Rated Power= 1.2kW; $T_{VSC}$= 0.04; $T_{L/C}$= 0.004.
\section*{Appendix B: System Parameters for Experimental Verification}

{\textit{Synchronous Machine}}: 415 $V_{L-L}$; 50 Hz; 3 kVA; 3-Phase; 1500 rpm.\\
\indent {\textit{Solar PV}} ($@ 1000 W/m^2$): $V_{oc}= 150.0 V, V_{mp}= 130.35 V, I_{sc}= 8.83 A,\\             I_{mp}= 8.41 A, P_{mp}= 1096.36W $.\\
\indent {\textit{DC-DC Boost Converter}}= 1.6 mH.\\
\indent {\textit{VSC Interfacing Inductance}}= 7.5 mH.\\ 
\indent {\textit{VSC Ripple Filter}}= 25 $\mu F$.\\ 
\indent {\textit{DC Link Voltage}}= 250 V.\\ 
\indent {\textit{DC Link Capacitance}}= 4500 $\mu F$.\\ 

\bibliography{mybibfile}

\begin{thebibliography}{10}
\expandafter\ifx\csname url\endcsname\relax
  \def\url#1{\texttt{#1}}\fi
\expandafter\ifx\csname urlprefix\endcsname\relax\def\urlprefix{URL }\fi
\expandafter\ifx\csname href\endcsname\relax
  \def\href#1#2{#2} \def\path#1{#1}\fi

\bibitem{REZAEI2015287}
N.~Rezaei, M.~Kalantar,
  \href{http://www.sciencedirect.com/science/article/pii/S0196890414010887}{Smart
  microgrid hierarchical frequency control ancillary service provision based on
  virtual inertia concept: An integrated demand response and droop controlled
  distributed generation framework}, Energy Conversion and Management 92 (2015)
  287 -- 301.
\newblock \href
  {http://dx.doi.org/https://doi.org/10.1016/j.enconman.2014.12.049}
  {\path{doi:https://doi.org/10.1016/j.enconman.2014.12.049}}.
\newline\urlprefix\url{http://www.sciencedirect.com/science/article/pii/S0196890414010887}

\bibitem{LIU2018169}
J.~Liu, M.~Hossain, J.~Lu, F.~Rafi, H.~Li,
  \href{http://www.sciencedirect.com/science/article/pii/S0378779618301524}{A
  hybrid ac/dc microgrid control system based on a virtual synchronous
  generator for smooth transient performances}, Electric Power Systems Research
  162 (2018) 169 -- 182.
\newblock \href {http://dx.doi.org/https://doi.org/10.1016/j.epsr.2018.05.014}
  {\path{doi:https://doi.org/10.1016/j.epsr.2018.05.014}}.
\newline\urlprefix\url{http://www.sciencedirect.com/science/article/pii/S0378779618301524}

\bibitem{486595}
H.~{Asano}, K.~{Yajima}, Y.~{Kaya}, Influence of photovoltaic power generation
  on required capacity for load frequency control, IEEE Transactions on Energy
  Conversion 11~(1) (1996) 188--193.

\bibitem{1597338}
A.~{Woyte}, V.~{Van Thong}, R.~{Belmans}, J.~{Nijs}, Voltage fluctuations on
  distribution level introduced by photovoltaic systems, IEEE Transactions on
  Energy Conversion 21~(1) (2006) 202--209.

\bibitem{5677458}
M.~{Datta}, T.~{Senjyu}, A.~{Yona}, T.~{Funabashi}, C.~{Kim}, A
  frequency-control approach by photovoltaic generator in a pv–diesel hybrid
  power system, IEEE Transactions on Energy Conversion 26~(2) (2011) 559--571.

\bibitem{8804864}
X.~{Ji}, Q.~{Liu}, Z.~{Liu}, Y.~{Xie}, J.~{Zhai}, Coordinated control and power
  management of diesel-pv-battery in hybrid stand-alone microgrid system, The
  Journal of Engineering 2019~(18) (2019) 5245--5249.

\bibitem{ROSINI2021106974}
A.~Rosini, D.~Mestriner, A.~Labella, A.~Bonfiglio, R.~Procopio,
  \href{https://www.sciencedirect.com/science/article/pii/S0378779620307720}{A
  decentralized approach for frequency and voltage regulation in islanded
  pv-storage microgrids}, Electric Power Systems Research 193 (2021) 106974.
\newblock \href {http://dx.doi.org/https://doi.org/10.1016/j.epsr.2020.106974}
  {\path{doi:https://doi.org/10.1016/j.epsr.2020.106974}}.
\newline\urlprefix\url{https://www.sciencedirect.com/science/article/pii/S0378779620307720}

\bibitem{HIRASE2018699}
Y.~Hirase, K.~Abe, K.~Sugimoto, K.~Sakimoto, H.~Bevrani, T.~Ise,
  \href{https://www.sciencedirect.com/science/article/pii/S0306261917308097}{A
  novel control approach for virtual synchronous generators to suppress
  frequency and voltage fluctuations in microgrids}, Applied Energy 210 (2018)
  699--710.
\newblock \href
  {http://dx.doi.org/https://doi.org/10.1016/j.apenergy.2017.06.058}
  {\path{doi:https://doi.org/10.1016/j.apenergy.2017.06.058}}.
\newline\urlprefix\url{https://www.sciencedirect.com/science/article/pii/S0306261917308097}

\bibitem{9113059}
A.~Kumar~V., S.~Sharma, A.~Verma, Optimal der sizing and dispatch of chp for a
  remote educational microgrid, in: 2020 IEEE 9th Power India International
  Conference (PIICON), 2020, pp. 1--6.
\newblock \href {http://dx.doi.org/10.1109/PIICON49524.2020.9113059}
  {\path{doi:10.1109/PIICON49524.2020.9113059}}.

\bibitem{AZIZ2022122458}
A.~S. Aziz, M.~F.~N. Tajuddin, M.~K. Hussain, M.~R. Adzman, N.~H. Ghazali,
  M.~A. Ramli, T.~E. {Khalil Zidane},
  \href{https://www.sciencedirect.com/science/article/pii/S0360544221027079}{A
  new optimization strategy for wind/diesel/battery hybrid energy system},
  Energy 239 (2022) 122458.
\newblock \href
  {http://dx.doi.org/https://doi.org/10.1016/j.energy.2021.122458}
  {\path{doi:https://doi.org/10.1016/j.energy.2021.122458}}.
\newline\urlprefix\url{https://www.sciencedirect.com/science/article/pii/S0360544221027079}

\bibitem{LIU2021125733}
B.~Liu, Z.~Wang, L.~Feng, K.~Jermsittiparsert,
  \href{https://www.sciencedirect.com/science/article/pii/S0959652620357796}{Optimal
  operation of photovoltaic/diesel generator/pumped water reservoir power
  system using modified manta ray optimization}, Journal of Cleaner Production
  289 (2021) 125733.
\newblock \href
  {http://dx.doi.org/https://doi.org/10.1016/j.jclepro.2020.125733}
  {\path{doi:https://doi.org/10.1016/j.jclepro.2020.125733}}.
\newline\urlprefix\url{https://www.sciencedirect.com/science/article/pii/S0959652620357796}

\bibitem{KUMAR2021102965}
P.~Kumar, N.~Pal, H.~Sharma,
  \href{https://www.sciencedirect.com/science/article/pii/S2352152X21006794}{Techno-economic
  analysis of solar photo-voltaic/diesel generator hybrid system using
  different energy storage technologies for isolated islands of india}, Journal
  of Energy Storage 41 (2021) 102965.
\newblock \href {http://dx.doi.org/https://doi.org/10.1016/j.est.2021.102965}
  {\path{doi:https://doi.org/10.1016/j.est.2021.102965}}.
\newline\urlprefix\url{https://www.sciencedirect.com/science/article/pii/S2352152X21006794}

\bibitem{MARQUSEE2020114918}
J.~Marqusee, D.~Jenket,
  \href{https://www.sciencedirect.com/science/article/pii/S030626192030430X}{Reliability
  of emergency and standby diesel generators: Impact on energy resiliency
  solutions}, Applied Energy 268 (2020) 114918.
\newblock \href
  {http://dx.doi.org/https://doi.org/10.1016/j.apenergy.2020.114918}
  {\path{doi:https://doi.org/10.1016/j.apenergy.2020.114918}}.
\newline\urlprefix\url{https://www.sciencedirect.com/science/article/pii/S030626192030430X}

\bibitem{MARQUSEE2021116437}
J.~Marqusee, S.~Ericson, D.~Jenket,
  \href{https://www.sciencedirect.com/science/article/pii/S0306261921000052}{Impact
  of emergency diesel generator reliability on microgrids and building-tied
  systems}, Applied Energy 285 (2021) 116437.
\newblock \href
  {http://dx.doi.org/https://doi.org/10.1016/j.apenergy.2021.116437}
  {\path{doi:https://doi.org/10.1016/j.apenergy.2021.116437}}.
\newline\urlprefix\url{https://www.sciencedirect.com/science/article/pii/S0306261921000052}

\bibitem{5648756}
A.~{Elmitwally}, M.~{Rashed}, Flexible operation strategy for an isolated
  pv-diesel microgrid without energy storage, IEEE Transactions on Energy
  Conversion 26~(1) (2011) 235--244.

\bibitem{RASHED2008949}
M.~Rashed, A.~Elmitwally, S.~Kaddah,
  \href{https://www.sciencedirect.com/science/article/pii/S0378779607001654}{New
  control approach for a pv-diesel autonomous power system}, Electric Power
  Systems Research 78~(6) (2008) 949--956.
\newblock \href {http://dx.doi.org/https://doi.org/10.1016/j.epsr.2007.07.003}
  {\path{doi:https://doi.org/10.1016/j.epsr.2007.07.003}}.
\newline\urlprefix\url{https://www.sciencedirect.com/science/article/pii/S0378779607001654}

\bibitem{8513881}
Y.~{Mi}, Y.~{Song}, Y.~{Fu}, X.~{Su}, C.~{Wang}, J.~{Wang}, Frequency and
  voltage coordinated control for isolated wind–diesel power system based on
  adaptive sliding mode and disturbance observer, IEEE Transactions on
  Sustainable Energy 10~(4) (2019) 2075--2083.

\bibitem{7478160}
C.~{Wang}, Y.~{Mi}, Y.~{Fu}, P.~{Wang}, Frequency control of an isolated
  micro-grid using double sliding mode controllers and disturbance observer,
  IEEE Transactions on Smart Grid 9~(2) (2018) 923--930.

\bibitem{8676274}
D.~{Sharma}, S.~{Mishra}, Disturbance-observer-based frequency regulation
  scheme for low-inertia microgrid systems, IEEE Systems Journal 14~(1) (2020)
  782--792.

\bibitem{9224611}
M.~S. Alam, F.~S. Al-Ismail, A.~Salem, M.~A. Abido, High-level penetration of
  renewable energy sources into grid utility: Challenges and solutions, IEEE
  Access 8 (2020) 190277--190299.
\newblock \href {http://dx.doi.org/10.1109/ACCESS.2020.3031481}
  {\path{doi:10.1109/ACCESS.2020.3031481}}.

\bibitem{8630730}
H.~Hua, C.~Hao, Y.~Qin, J.~Cao, Stochastic robust h$_\infty$ control strategy
  for coordinated frequency regulation in energy internet considering time
  delay and uncertainty, in: 2018 13th World Congress on Intelligent Control
  and Automation (WCICA), 2018, pp. 111--118.
\newblock \href {http://dx.doi.org/10.1109/WCICA.2018.8630730}
  {\path{doi:10.1109/WCICA.2018.8630730}}.

\bibitem{7764193}
L.~Luo, X.~Zhao, X.~Li, W.~Yan, G.~Liu, P.~Zhou, L.~Wen, Effects of
  uncertainties in frequency regulations on probabilistic power flow analysis,
  in: 2016 International Conference on Probabilistic Methods Applied to Power
  Systems (PMAPS), 2016, pp. 1--6.
\newblock \href {http://dx.doi.org/10.1109/PMAPS.2016.7764193}
  {\path{doi:10.1109/PMAPS.2016.7764193}}.

\bibitem{Uncertainty}
C.~Yan, T.~Yi, J.~Dai, C.~Wang, S.~Wu, Uncertainty modeling of wind power
  frequency regulation potential considering distributed characteristics of
  forecast errors, Protection and Control of Modern Power Systems 6.
\newblock \href {http://dx.doi.org/10.1186/s41601-021-00200-3}
  {\path{doi:10.1186/s41601-021-00200-3}}.

\bibitem{ABEDINI2019100200}
M.~Abedini, E.~Mahmodi, M.~Mousavi, I.~Chaharmahali,
  \href{https://www.sciencedirect.com/science/article/pii/S2352467718303485}{A
  novel fuzzy pi controller for improving autonomous network by considering
  uncertainty}, Sustainable Energy, Grids and Networks 18 (2019) 100200.
\newblock \href {http://dx.doi.org/https://doi.org/10.1016/j.segan.2019.100200}
  {\path{doi:https://doi.org/10.1016/j.segan.2019.100200}}.
\newline\urlprefix\url{https://www.sciencedirect.com/science/article/pii/S2352467718303485}

\bibitem{LANKESHWARA2022117971}
G.~Lankeshwara, R.~Sharma, R.~Yan, T.~K. Saha,
  \href{https://www.sciencedirect.com/science/article/pii/S0306261921012757}{Control
  algorithms to mitigate the effect of uncertainties in residential demand
  management}, Applied Energy 306 (2022) 117971.
\newblock \href
  {http://dx.doi.org/https://doi.org/10.1016/j.apenergy.2021.117971}
  {\path{doi:https://doi.org/10.1016/j.apenergy.2021.117971}}.
\newline\urlprefix\url{https://www.sciencedirect.com/science/article/pii/S0306261921012757}

\bibitem{8283546}
N.~Nguyen-Hong, H.~Nguyen-Duc, Y.~Nakanishi, Optimal sizing of energy storage
  devices in isolated wind-diesel systems considering load growth uncertainty,
  IEEE Transactions on Industry Applications 54~(3) (2018) 1983--1991.
\newblock \href {http://dx.doi.org/10.1109/TIA.2018.2802940}
  {\path{doi:10.1109/TIA.2018.2802940}}.

\bibitem{6894151}
N.~Mendis, K.~M. Muttaqi, S.~Perera, S.~Kamalasadan, An effective power
  management strategy for a wind–diesel–hydrogen-based remote area power
  supply system to meet fluctuating demands under generation uncertainty, IEEE
  Transactions on Industry Applications 51~(2) (2015) 1228--1238.
\newblock \href {http://dx.doi.org/10.1109/TIA.2014.2356013}
  {\path{doi:10.1109/TIA.2014.2356013}}.

\bibitem{314514}
K.~Vidyanandan, N.~Senroy,
  \href{https://ietresearch.onlinelibrary.wiley.com/doi/abs/10.1049/iet-gtd.2015.0449}{Frequency
  regulation in a wind–diesel powered microgrid using flywheels and fuel
  cells}, IET Generation, Transmission \& Distribution 10~(3) (2016) 780--788.
\newblock \href
  {http://arxiv.org/abs/https://ietresearch.onlinelibrary.wiley.com/doi/pdf/10.1049/iet-gtd.2015.0449}
  {\path{arXiv:https://ietresearch.onlinelibrary.wiley.com/doi/pdf/10.1049/iet-gtd.2015.0449}},
  \href {http://dx.doi.org/https://doi.org/10.1049/iet-gtd.2015.0449}
  {\path{doi:https://doi.org/10.1049/iet-gtd.2015.0449}}.
\newline\urlprefix\url{https://ietresearch.onlinelibrary.wiley.com/doi/abs/10.1049/iet-gtd.2015.0449}

\bibitem{ABUBAKR2021106814}
H.~Abubakr, T.~H. Mohamed, M.~M. Hussein, J.~M. Guerrero, G.~Agundis-Tinajero,
  \href{https://www.sciencedirect.com/science/article/pii/S0142061521000545}{Adaptive
  frequency regulation strategy in multi-area microgrids including renewable
  energy and electric vehicles supported by virtual inertia}, International
  Journal of Electrical Power \& Energy Systems, 129 (2021) 106814.
\newblock \href
  {http://dx.doi.org/https://doi.org/10.1016/j.ijepes.2021.106814}
  {\path{doi:https://doi.org/10.1016/j.ijepes.2021.106814}}.
\newline\urlprefix\url{https://www.sciencedirect.com/science/article/pii/S0142061521000545}

\bibitem{RAHMAN2016488}
M.~S. Rahman, M.~Hossain, J.~Lu,
  \href{https://www.sciencedirect.com/science/article/pii/S0196890416304484}{Coordinated
  control of three-phase ac and dc type ev–esss for efficient hybrid
  microgrid operations}, Energy Conversion and Management 122 (2016) 488--503.
\newblock \href
  {http://dx.doi.org/https://doi.org/10.1016/j.enconman.2016.05.070}
  {\path{doi:https://doi.org/10.1016/j.enconman.2016.05.070}}.
\newline\urlprefix\url{https://www.sciencedirect.com/science/article/pii/S0196890416304484}

\bibitem{8649677}
S.~{Sharma}, Y.~{Xu}, A.~{Verma}, B.~K. {Panigrahi}, Time-coordinated
  multienergy management of smart buildings under uncertainties, IEEE
  Transactions on Industrial Informatics 15~(8) (2019) 4788--4798.

\bibitem{6565075}
Y.~{Mi}, Y.~{Fu}, J.~B. {Zhao}, P.~{Wang}, The novel frequency control method
  for pv-diesel hybrid system, in: 2013 10th IEEE International Conference on
  Control and Automation (ICCA), 2013, pp. 180--183.

\bibitem{osti_5599996}
O.~I. Elgerd, Electric energy systems theory: an introduction.

\bibitem{10.5555/2636749}
S.~Li, J.~Yang, W.-H. Chen, X.~Chen, Disturbance Observer-Based Control:
  Methods and Applications, 1st Edition, CRC Press, Inc., USA, 2014.

\bibitem{8398456}
A.~Fathi, Q.~Shafiee, H.~Bevrani, Robust frequency control of microgrids using
  an extended virtual synchronous generator, IEEE Transactions on Power Systems
  33~(6) (2018) 6289--6297.
\newblock \href {http://dx.doi.org/10.1109/TPWRS.2018.2850880}
  {\path{doi:10.1109/TPWRS.2018.2850880}}.

\bibitem{SOLIMAN2021107216}
M.~Soliman, M.~Ali,
  \href{https://www.sciencedirect.com/science/article/pii/S0142061521004555}{Parameterization
  of robust multi-objective pid-based automatic voltage regulators: Generalized
  hurwitz approach}, International Journal of Electrical Power \& Energy
  Systems 133 (2021) 107216.
\newblock \href
  {http://dx.doi.org/https://doi.org/10.1016/j.ijepes.2021.107216}
  {\path{doi:https://doi.org/10.1016/j.ijepes.2021.107216}}.
\newline\urlprefix\url{https://www.sciencedirect.com/science/article/pii/S0142061521004555}

\end{thebibliography}
\end{document}